# Nucleon Properties at Finite Temperature in the Extended Quark-Sigma Model

**M. Abu-Shady* and A. K. Abu-Nab**

Department of Mathematics, Faculty of Science, Menoufia University, Egypt

**Email address:**
abu_shady_1999@yahoo.com (M. Abu-Shady)

**Abstract:** Hadron properties are studied at hot medium using the quark sigma model. The quark sigma model is extended to include eighth-order of mesonic interactions based on some aspects of quantum chromodynamic (QCD) theory. The extended effective potential tends to the original effective potential when the coupling between the higher order mesonic interactions equal to zero. The field equations have been solved in the mean-field approximation by using the extended iteration method. We found that the nucleon mass increases with increasing temperature and the magnetic moments of proton and neutron increase with increasing temperature. A comparison is presented with recent previous works and other models. We conclude that higher-order mesonic interactions play an important role in changing the behavior of nucleon properties at finite temperature. In addition, the deconfinement phase transition is satisfied in the present model.

## 1. Introduction

The Quantum Chromodynamics (QCD) theory is acceptable theory for describing strong interactions [1]. The lattice QCD is acceptable technique to solve the field equations of QCD theory. Unfortunately, the lattice QCD has difficulties at finite baryonic chemical potential [2]. Therefore, the models are applied, in which incorporate the spontaneous chiral symmetry of QCD such as the linear sigma model [3] and the Nambu-Jona-Laisino model [4]. These models provide a good description for hadron properties at zero and finite temperatures. The linear sigma model successfully predicts the nucleon properties at low energy after replacing the baryon picture by the quark picture as in Refs. [5-7].

In recent years, the model is extended to include higher-order mesonic interactions as in Refs. [9-11] to provide a good description of hadron properties at zero temperature. In addition, the model is extended to finite temperature with higher-order mesonic interactions as in Refs. [12-14]. In the same direction, the model is extended to include the ployakov loop which plays important role for description the critical temperature and phase transition as in Refs. [15-16].

The aim of this work is to study the effect of finite temperature on nucleon properties such as the nucleon mass, the magnetic moments of proton and neutron. In the following section, we briefly review the linear sigma model. The higher-order mesonic interactions are studied in Sec. 3. The numerical calculations and the discussion of results are presented in Secs. 4 and 5.

## 2. The Chiral-Quark Sigma Model

Birse and Banerjee [5] described the interactions of quarks via the exchange of $\sigma$ and $\pi$- meson fields. The Lagrangian density is

$$L(r) = i\bar{\psi}\partial_\mu\gamma^\mu\psi + \frac{1}{2}\left(\partial_\mu\sigma\partial^\mu\sigma + \partial_\mu\boldsymbol{\pi}\partial^\mu\boldsymbol{\pi}\right) + g\bar{\psi}(\sigma + i\gamma_5\tau.\boldsymbol{\pi})\psi - U_1^{(0)}(\sigma,\boldsymbol{\pi}), \quad (1)$$

with

$$U_1^{(0)}(\sigma,\pi) = \frac{\lambda^2}{4}(\sigma^2 + \boldsymbol{\pi}^2 - \upsilon^2) + m_\pi^2 f_\pi \sigma, \quad (2)$$

is the meson-meson interaction potential where the $\psi, \sigma$ and $\pi$ are the quark, sigma, and pion fields, respectively.

In the mean-field approximation, the meson fields treat as time-independent classical fields. This means that we replace the power and the products of the meson fields by the corresponding powers and the products of their expectation values. In Eq. (2), the meson-meson interaction leads to the hidden chiral symmetry $SU(2)\times SU(2)$ with $\sigma(r)$ taking on a vacuum expectation value

$$\langle\sigma\rangle = -f_\pi, \tag{3}$$

where $f_\pi = 93$ MeV is the pion decay constant. The final term in Eq. (2) is included to break the chiral symmetry explicitly. It leads to the partial conservation of

$$\lambda^2 = \frac{m_\sigma^2 - m_\pi^2}{2f_\pi^2}, \tag{4}$$

$$\nu^2 = f_\pi^2 - \frac{m_\pi^2}{\lambda^2}. \tag{5}$$

## 3. The Chiral Higher-Order Quark Sigma Model at Finite Temperature

The Lagrangian density of the extended linear sigma model which describes the interactions between quarks via the $\sigma$ and $\pi$- mesons [11],

$$L(r) = i\bar{\psi}\partial_\mu\gamma^\mu\psi + \frac{1}{2}\left(\partial_\mu\sigma\partial^\mu\sigma + \partial_\mu\boldsymbol{\pi}\partial^\mu\boldsymbol{\pi}\right) + g\bar{\psi}(\sigma + i\gamma_5\tau.\boldsymbol{\pi})\psi - U_2^{(0)}(\sigma,\boldsymbol{\pi}), \tag{6}$$

with

$$U_2^{(0)}(\sigma,\boldsymbol{\pi}) = \frac{\lambda_1^2}{4}(\sigma^2 + \boldsymbol{\pi}^2 - \nu_1^2)^2 + \frac{\lambda_2^2}{4}[(\sigma^2 + \boldsymbol{\pi}^2)^2 - \nu_2^2]^2 + m_\pi^2 f_\pi \sigma. \tag{7}$$

It is clear that potential satisfies the chiral symmetry when $m_\pi \to 0$ . In the original model [5], the higher-order term in Eq. 7 is excluded by the requirement of renormalizability. Since we are going to use Eq. (7) as an approximating effective model. Therefore, the model did not need and should not be renormalizable as in Ref. [17]. By using the PCAC and the minimization conditions of mesonic potential as in Ref. [11], we obtain

$$\lambda_1^2 = \frac{m_\sigma^2 - m_\pi^2}{4f_\pi^2} \quad , \quad \nu_1^2 = f_\pi^2 - \frac{m_\pi^2}{2\lambda_1^2}, \tag{8}$$

$$\lambda_2^2 = \frac{m_\sigma^2 - 3m_\pi^2}{16 f_\pi^6} \quad , \quad \nu_2^2 = f_\pi^4 - \frac{m_\pi^2}{2 f_\pi^2 \lambda_2^2}. \tag{9}$$

We can write the effective potential of the mesonic interaction at finite temperature [13].

$$U_2^{eff.}(\sigma,\boldsymbol{\pi},T) = U_2^{(0)}(\sigma,\boldsymbol{\pi}) + \frac{70\,\pi^2 T^4}{90} + \left(\frac{m_\sigma^2 - m_\pi^2}{24 f_\pi^2}\right)T^2 \times (\sigma^2 + \boldsymbol{\pi}^2) + \left(\frac{m_\sigma^2 - m_\pi^2}{24 f_\pi^2}\right)T^2\left(\sigma^2 + \boldsymbol{\pi}^2 - \frac{\nu_1^2}{2}\right). \tag{10}$$

Now we can expand the extremum with the shifted field defined as

$$\sigma = \acute{\sigma} - \sigma(T), \tag{11}$$

where

$$\sigma(T) = f_\pi(1 + \delta(T)),$$

and

$$\delta(T) = \frac{-\lambda_1^2 f_\pi^2 + \lambda_1^2\nu_1^2 - \lambda_2^2 f_\pi^6 + \lambda_2^2\nu_2^2 f_\pi^2 - m_\pi^2 - \frac{m_\sigma^2 - m_\pi^2}{6f_\pi^2}T^2}{\lambda_1^2 f_\pi^2 + \lambda_1^2\nu_1^2 + 10\lambda_2^2 f_\pi^6 - 2\lambda_2^2\nu_2^2 f_\pi^2 - 2m_\pi^2 - \frac{m_\sigma^2 - m_\pi^2}{6f_\pi^2}T^2}, \tag{12}$$

is defined in Ref. [13], substituting by Eq. 11 into Eq. 10, we get

$$U_2^{eff.}(\acute{\sigma},\boldsymbol{\pi},T) = \frac{\lambda_1^2}{4}((\acute{\sigma} - \sigma(T))^2 + \boldsymbol{\pi}^2 - \nu_1^2)^2 + \frac{\lambda_2^2}{4}(((\acute{\sigma} - \sigma(T))^2 + \boldsymbol{\pi}^2)^2 - \nu_2^2)^2 + m_\pi^2 f_\pi\ (\acute{\sigma} - \sigma(T)) + \frac{70\,\pi^2T^4}{90} + \left(\frac{m_\sigma^2 - m_\pi^2}{24{f_\pi}^2}\right)T^2((\acute{\sigma} - \sigma(T))^2 + \boldsymbol{\pi}^2) + \left(\frac{m_\sigma^2 - m_\pi^2}{24{f_\pi}^2}\right)T^2\left((\acute{\sigma} - \sigma(T))^2 + \boldsymbol{\pi}^2 - \frac{\nu^2}{2}\right). \tag{13}$$

The time-independent fields and satisfy the Euler Lagrange equations, and the quark wave function satisfies the Dirac eigen value equation. Substituting Eqs. (6 and 13) at finite temperature in Euler Lagrange equation, we get

$$\acute{\sigma} = g\bar{\psi}\,\psi - \lambda_1^2\left[((\acute{\sigma} - \sigma(T))^2 + \boldsymbol{\pi}^2 - \nu_1^2)(\acute{\sigma} - \sigma(T))\ \right] - -2\lambda_2^2\,(((\acute{\sigma} - \sigma(T))^2 + \boldsymbol{\pi}^2)^2 - \nu_2^2) \times ((\acute{\sigma} - \sigma(T))^2 + \boldsymbol{\pi}^2) \times (\acute{\sigma} - \sigma(T)) + \left(\frac{m_\sigma^2 - m_\pi^2}{6{f_\pi}^2}\right)(\acute{\sigma} - \sigma(T))T^2 + m_\pi^2 f_\pi\ \sigma(T) \tag{14}$$

$$\Box\pi = ig\bar{\psi}\,\gamma_5.\tau\,\psi - \lambda_1^2\left[((\acute{\sigma} - \sigma(T))^2 + \boldsymbol{\pi}^2 - \nu_1^2)\pi\ \right] - 2\lambda_2^2\,(((\acute{\sigma} - \sigma(T))^2 + \pi^2)^2 - \nu_2^2) \times ((\acute{\sigma} - \sigma(T))^2 + \boldsymbol{\pi}^2) \times \pi + \left(\frac{m_\sigma^2 - m_\pi^2}{6{f_\pi}^2}\right)\pi T^2 \tag{15}$$

where $\tau$ refers to Pauli isospin matrices, $\gamma_5 = \begin{pmatrix}0 & 1\\ 1 & 0\end{pmatrix}$. Including the color force of freedom, one has $g\bar{\psi}\psi \to N_C\, g\bar{\psi}\psi$ where $N_C = 3$ colors. Thus

$$\psi(r) = \frac{1}{\sqrt{4\pi}}\begin{bmatrix}u(r)\\ iw(r)\end{bmatrix} \text{ and } \bar{\psi}(r) = \frac{1}{\sqrt{4\pi}}[u(r)\quad iw(r)], \tag{16}$$

then,

$$\rho_s = N_C\,\bar{\psi}\psi = \frac{3g}{4\pi}(u^2 - w^2), \tag{17}$$

$$\rho_p = iN_C\,\bar{\psi}\ \boldsymbol{\gamma_5}.\boldsymbol{\tau}\psi = \frac{3g}{2\pi}(uw), \tag{18}$$

$$\rho_v = \frac{3g}{4\pi}(u^2 + w^2), \tag{19}$$

where $\rho_s$, $\rho_p$ and $\rho_v$ are sigma, pion and vector densities, respectively. These equations are subject to the boundary conditions as follows,

$$\sigma(r) \sim -\sigma(T), \qquad \pi(r) \sim 0 \quad \text{at} \quad r \to \infty. \tag{20}$$

by using hedgehog ansatz [8], where

$$\boldsymbol{\pi}(r) = \pi(r)\hat{\boldsymbol{r}}. \tag{21}$$

The chiral Dirac equation for the quark is [8]

$$\frac{du}{dr} = -P(r)u + \left(W - m_q + S(r)\right) w, \qquad (22)$$

where the scalar potential $S(r) = g\langle \acute{\sigma} \rangle$, the pseudoscalar potential $P(r) = \langle \pi . \hat{r} \rangle$, and W is the eigen-value of quarks spinor $\psi$

$$\frac{dw}{dr} = -\left(W - m_q + S(r)\right) u - \left(\frac{2}{r} - P(r)\right). \qquad (23)$$

# 4. Numerical Calculations and Discussions

## 4.1 The Scalar Field $\acute{\sigma}$

To solve Eq. (14), we integrate a suitable Green's function over the source fields as in Refs. [8, 9]. Thus

$$\acute{\sigma}(r) = m_\sigma \int_0^\infty \acute{r}^2 d\acute{r} \left(\frac{sinh(m_\sigma r <) exp(-m_\sigma r >)}{m_\sigma r >}\right)$$

$$\{g\, \rho_s(\acute{r}) - \lambda_1^2 \left[\left((\acute{\sigma} - \sigma(T))^2 + \boldsymbol{\pi}^2 - v_1^2\right)(\acute{\sigma} - \sigma(T))\right]$$

$$-2\lambda_2^2 \left(((\acute{\sigma} - \sigma(T))^2 + \boldsymbol{\pi}^2)^2 - v_2^2\right) \times \left((\acute{\sigma} - \sigma(T))^2 + \boldsymbol{\pi}^2\right) \times (\acute{\sigma} - \sigma(T)) + \quad +$$

$$\left(\frac{m_\sigma^2 - m_\pi^2}{6 f_\pi{}^2}\right)(\acute{\sigma} - \sigma(T)) T^2 + m_\pi^2 f_\pi\ \sigma(T)\}. \qquad (24)$$

Note that this form is implicit in the solution of $\acute{\sigma}$ involves integrals over the unknown $\acute{\sigma}$ itself. We will solve this implicit integral equation by iterating to self-consistency.

## 4.1 The pion field $\boldsymbol{\pi}$

To solve Eq. (15), we integrate a suitable Green's function over the source fields.

$$\pi(r) = m_\pi \int_0^\infty \acute{r}^2\, d\acute{r} \left(\frac{[-sinh(m_\pi r<) + m_\pi r< \cosh(m_\pi r<)]}{(m_\pi r>)^2}\right)$$

$$\times \{g\, \rho_p(\acute{r}) - \lambda_1^2 [((\acute{\sigma} - \sigma(T))^2 + \boldsymbol{\pi}^2 - v_1^2)\pi \ ]$$

$$-2\lambda_2^2 \left(((\acute{\sigma} - \sigma(T))^2 + \pi^2)^2 - v_2^2\right) \times \left((\acute{\sigma} - \sigma(T))^2 + \boldsymbol{\pi}^2\right) \times \quad \times \pi$$

$$+ \left(\frac{m_\sigma^2 - m_\pi^2}{6 f_\pi{}^2}\right) \pi T^2\} \qquad (25)$$

We have solved Dirac Eqs. (22), (23) using fourth-order Rung Kutta method. Due to the implicit nonlinearly of these Esq. (24), (25) it is necessary to iterate the solution until self-consistency is achieved. To start this iteration process, we could use the chiral circle form for the meson fields [8, 9]:

$$S(r) = m_q(1 - \cos\theta) \quad , \quad P(r) = -m_q \sin\theta, \qquad (26)$$

# 5. Discussion of the Results

In present work, we investigate hadron properties such as the hedgehog mass, the magnetic moments of proton and neutron and the pion-nucleon coupling constant in the linear sigma model, in which the higher-order mesonic interactions up to the eighth-order interactions are included. We study the case with two massless quark flavors (u, d) and $N_c = 3$, which is the number of colors. We solved the field equations in the mean-field approximation by using the extended iteration method described in the section 3. For numerical computation, we use the free parameters at zero temperature as the initial conditions, namely we took $m_\sigma = 500 \to 700$ MeV, $f_\pi = 93$ MeV and the coupling constant g as a free parameter. The results present in the figures (1, 2, 3, 4 and 5) and Table (1). In Fig. (1), the hedgehog mass is plotted as a function of temperature for different values of sigma mass. At zero temperature, the hedgehog mass equal to 1051 MeV at $m_\sigma = 600$ MeV. The obtained value of hedgehog mass is in good agreement with experimental data equal to $M_H$ =1086 MeV with relative error about 3%. By increasing temperature, the hedgehog mass increases, which indicates to the deconfinement chiral phase transition is satisfied in the present work. This result is in agreement with Ref. [2]. By increasing sigma mass, we note that the curves shift to higher values. The effect of sigma mass appears at higher-values of temperatures. This effected is not calculated in Ref. [14]. In Fig. (2), the magnetic moment of proton is plotted as a function of temperature. We note that the magnetic of proton increases with increasing temperature. This result is in good agreement with Chirstov et al. [1]. They found that the magnetic of proton increases with increasing temperature using the same model without including higher-mesonic interactions. A similar result is obtained with respect to magnetic moment of neutron, where the magnetic moment of neutron inecreases with increasing temperature as in Fig. (3). In Fig. (4), the coupling constant $g_A(T)$ is plotted as a function of temperature. We note that the axial coupling constant $g_A(T)$ decreases with increasing temperature. The curve shifts to lower values by increasing sigma mass. A similar behaviour is found for the pion-nucleon coupling constant $\boldsymbol{g_{\pi NN}}$. The pion-nucleon coupling decreases with increasing temperature. In comparison with Ref. [14], the authors found that the nucleon–coupling constant is a monotonically decreasing function of temperature (T). This indicates a deconfinement phase transition is satisfied in the present work.

In Table (1), we show the dynamics of the hedgehog mass at finite temperature (T=80 MeV). We note that the kinetic energies of quark, sigma and pion fields increase with increasing temperature. In addition, pion interaction energy and meson interaction energy are more affected by increasing temperature. Therefore , the total of energies give the hedgehog mass that increases with increasing temperature.

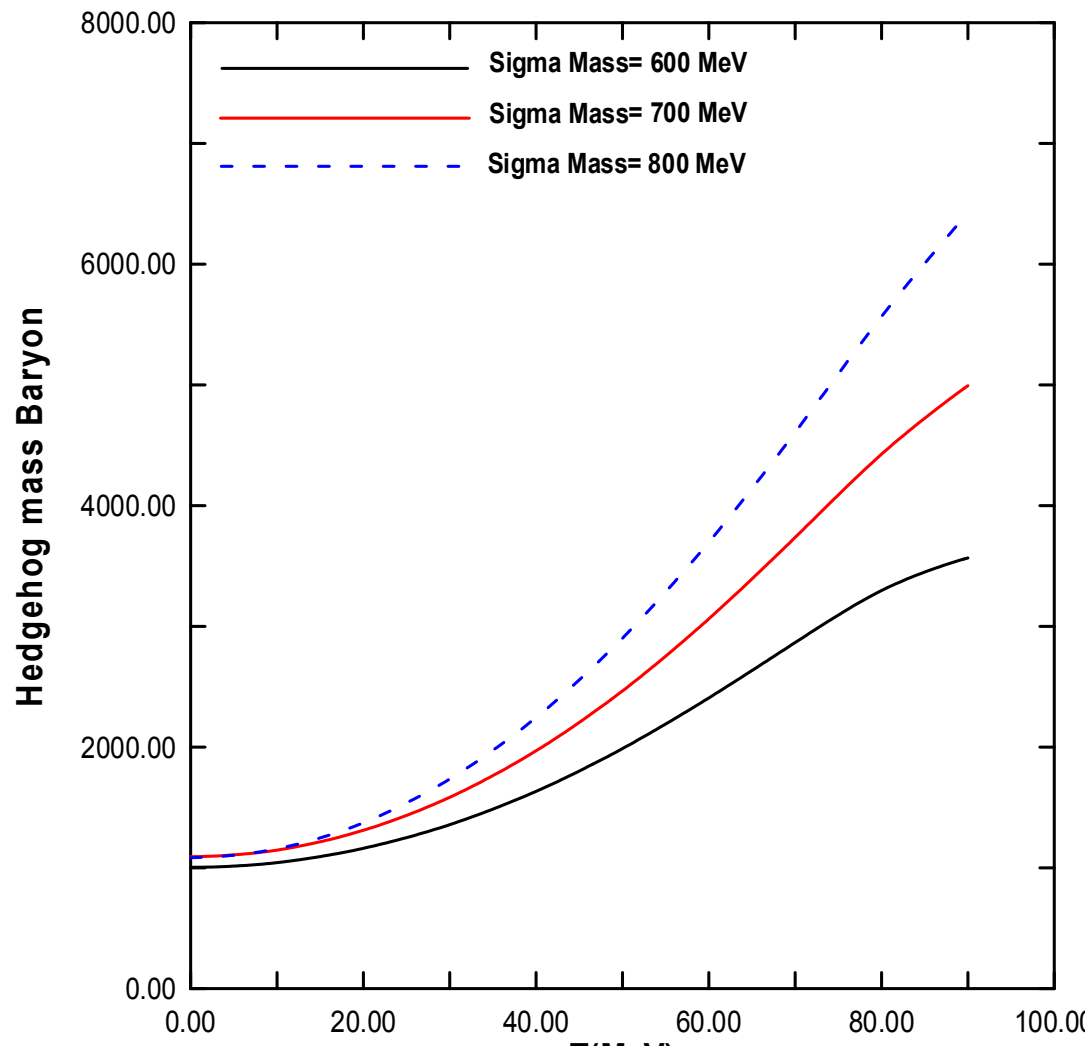

***Fig.1.*** *Hedgehog mass is plotted as a function of temperature at different values of the sigma masses*

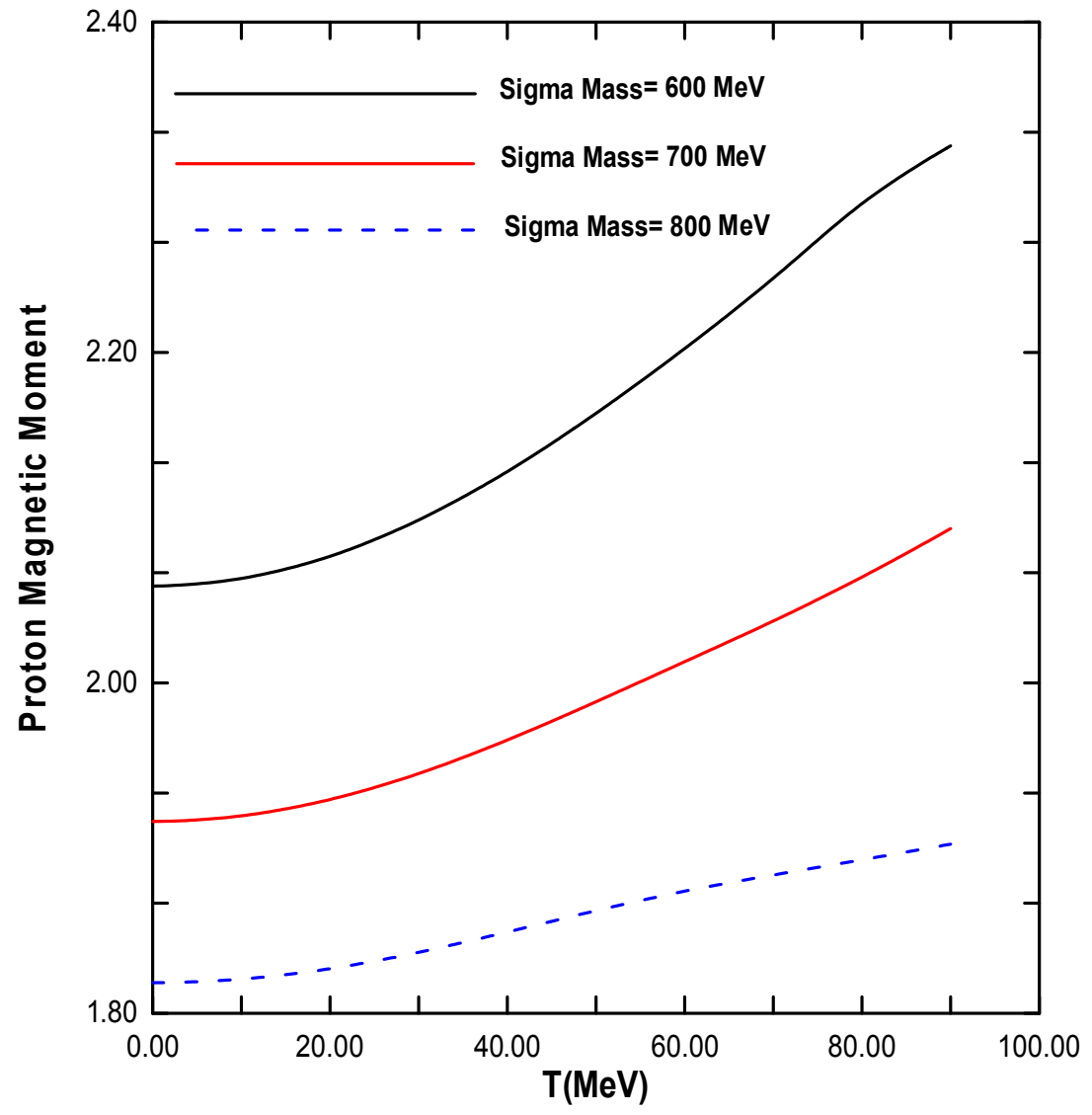

***Fig.2.*** *The proton magnetic moment is plotted as a function of temperature at different values of the sigma masses.*

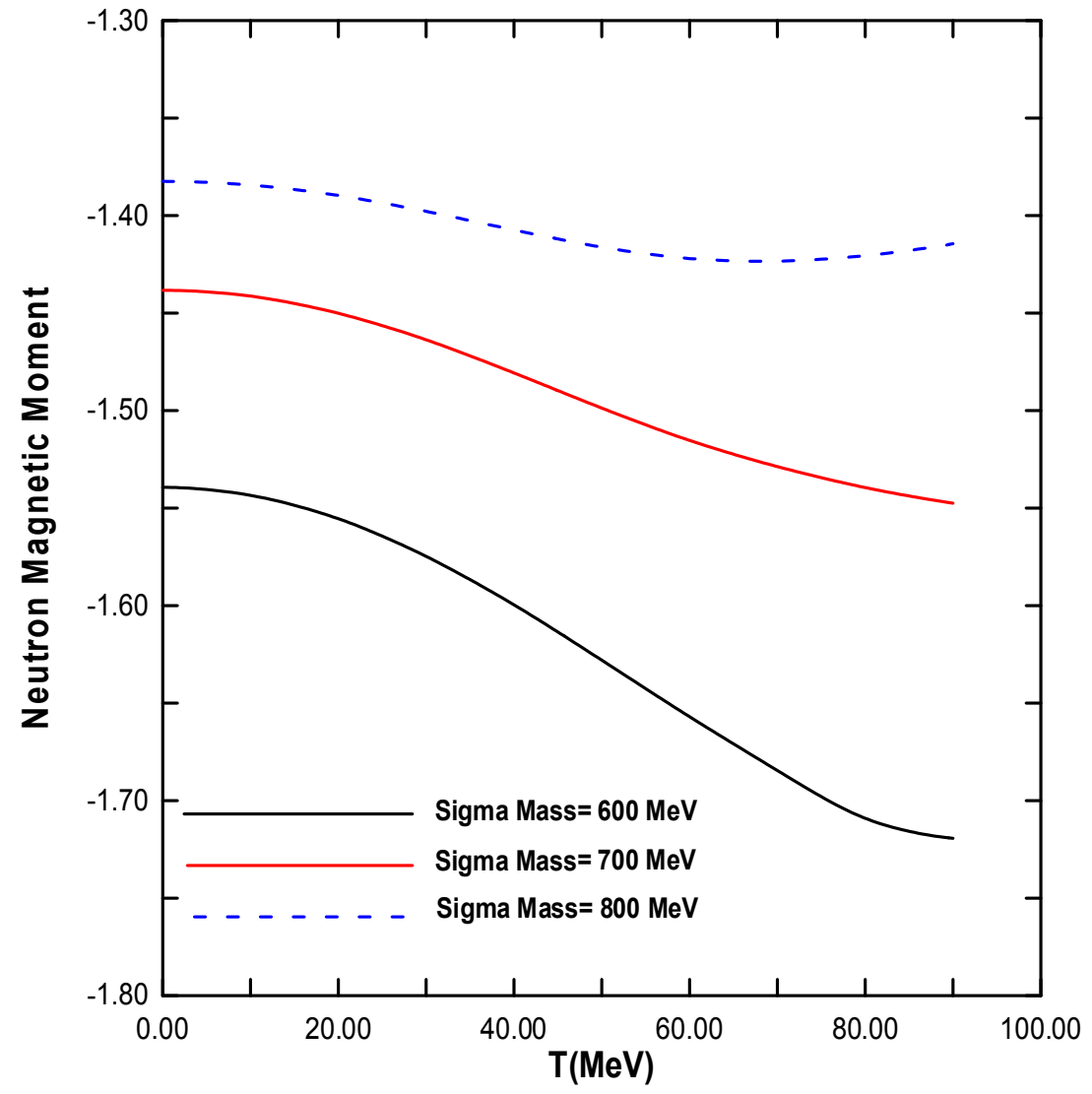

***Fig.3****. The neutron magnetic moment is plotted as a function of temperature at different values of the sigma masses.*

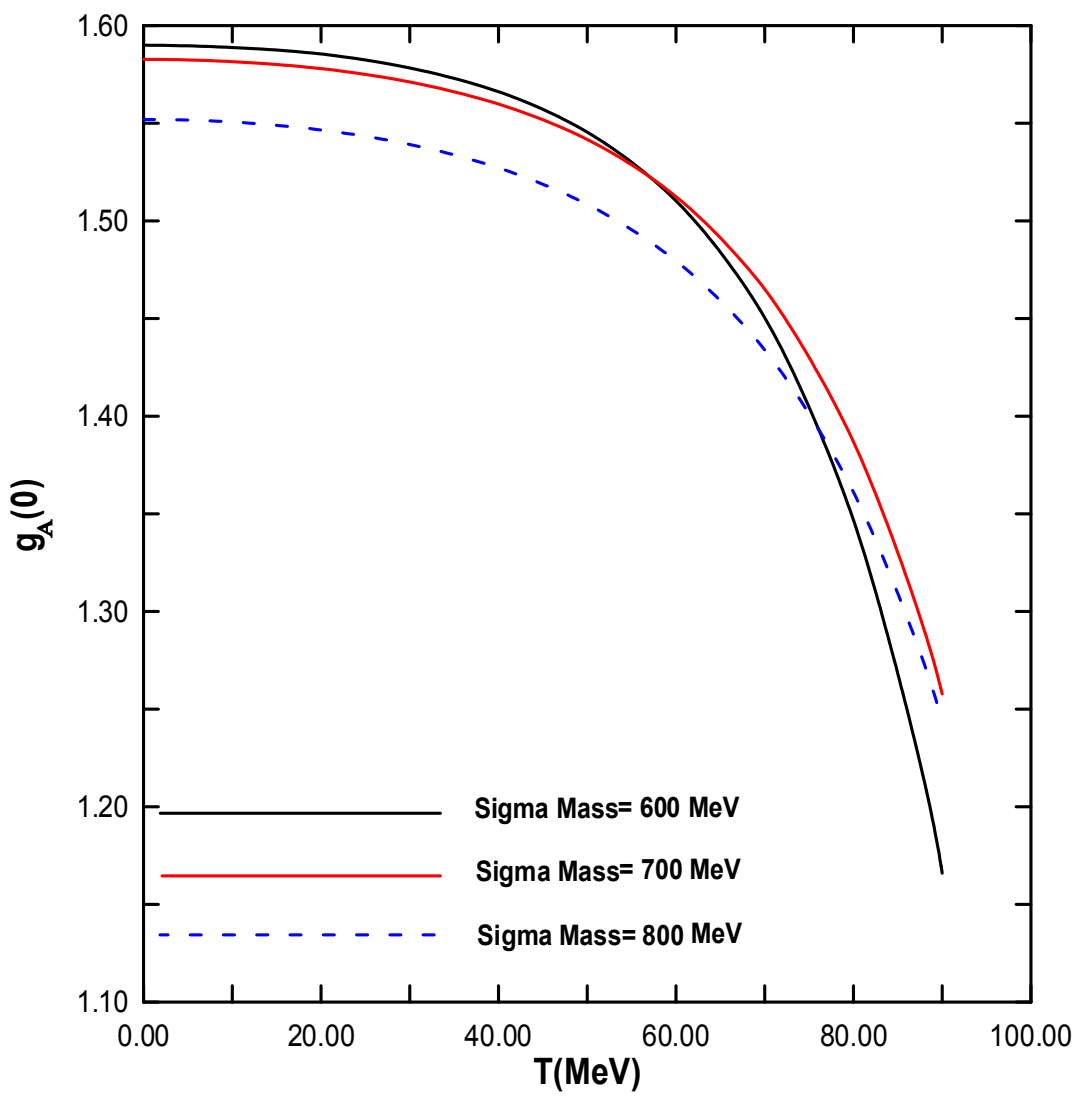

***Fig.4.*** *The axial coupling constant* $g_A(T)$ *is plotted as a function of temperature of different values o f sigma masses.*

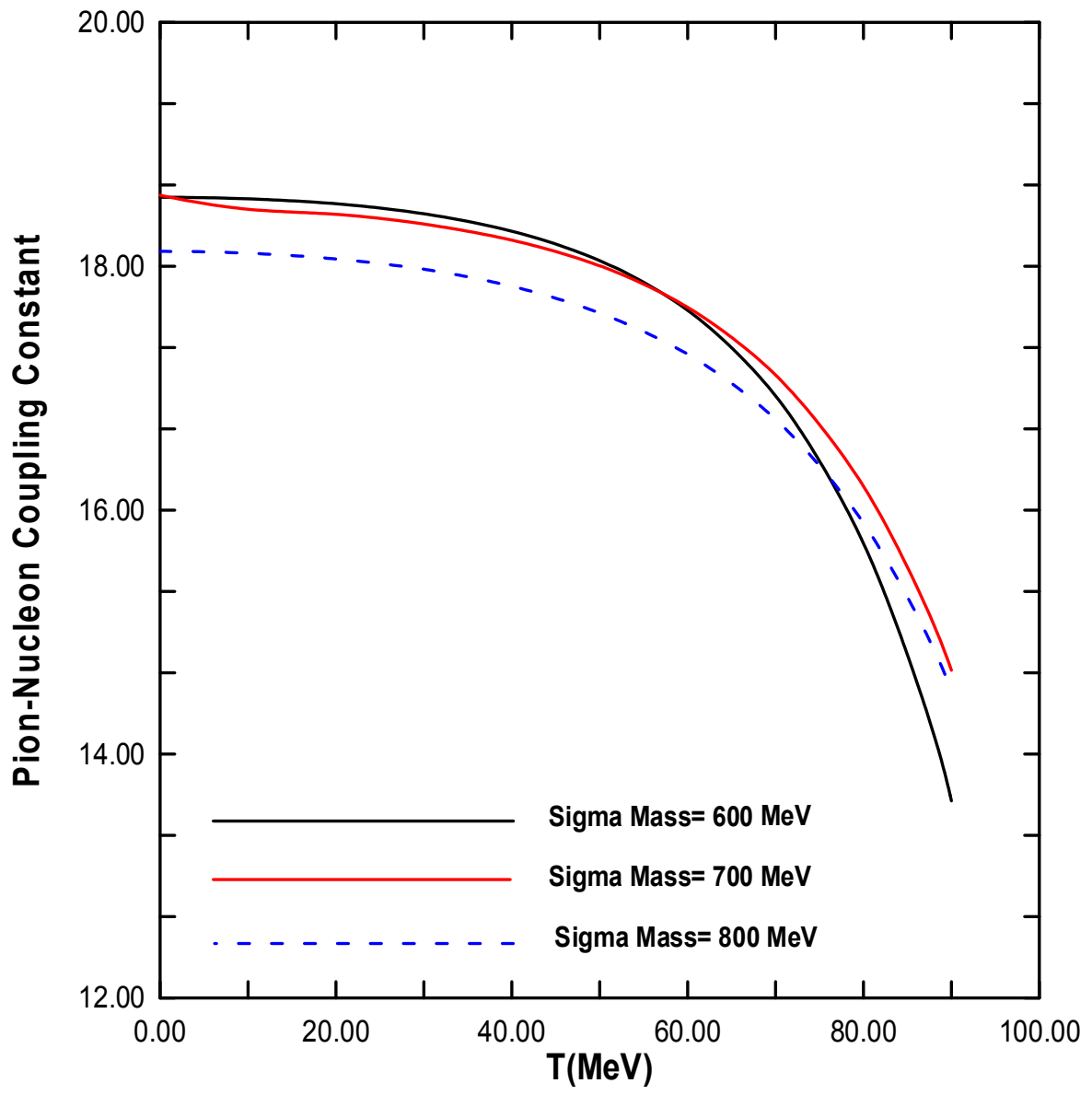

***Fig.5****. The pion-nucleon coupling constant* $g_{\pi NN}$ *is plotted as a function of temperature of different values o f sigma masses.*

***Table (1)****. Details of energy calculations of the hedgehog mass at* $m_\pi = 139.6$ *MeV,* $m_\sigma = 600$ *MeV, and* $m_q = 400$ *MeV. All quantities in MeV.*

| **Quantity** | **Temperature (T=0)** | **Temperature (T=80)** |
|---|---|---|
| Quark kinetic energy | 831.470 | 1308.365 |
| Sigma kinetic energy | 119.232 | 166.193 |
| Pion kinetic energy | 191.617 | 385.824 |
| Sigma-quark interaction energy | 405.425 | -182.980 |
| Pion-quark interaction energy | -378.244 | -954.867 |
| Meson-Meson interaction | -117.552 | 1751.056 |
| Hedgehog mass | 1051.9 | 2473.7 |

## 6. Comparison with other works

It is interesting to compare the present results with the results from other works. The higher-order mesonic and quark interactions are taken in the quark models. In this comparison, we will concentrate on the two models, the chiral quark sigma model and the Nambu-Jona-Lasino model. The nucleon properties at finite temperature are studied by Christov et al. [1] in the framework of the quark sigma model, the qualitative behavior of nucleon properties such as the nucleon mass and the magnetic moment of proton and neutron is in agreement with the present results. In the work of Christov et al. [1], the higher-order mesonic interactions are not taken in their calculations. In addition, they used form factors to calculate the nucleon properties. In the original sigma model as in Ref. [18]. The nucleon properties are investigated at finite temperature without taken the higher-order mesonic interactions in the model. Abu-shady et al. [12, 13] consider the higher-order mesonic interactions in the quark sigma model by using different versions of effective potential. They study the effect of higher-order mesonic interactions on the chiral phase transition and the critical temperature, and meson masses. Therefore the nucleon properties are not calculated in Refs. [12, 13]. In comparison with Ref. [14], the author takes the higher-order mesonic interactions in the original sigma model and nucleon properties at finite temperature are calculated. The qualitative features of nucleon properties are in agreement with the present results as in Fig. 6. We have some advantages in comparison with Ref. [14]. The first, we can obtain the original meson-meson potential when the coupling of higher-order interactions is vanished as in Eq 7. This advantage is not found in Ref. [14]. The second, the effect of sigma mass on nucleon properties is not calculated in Ref. [14] which has strong effect on nucleon properties as discussed in the above section. The third, we expanded the sigma field around $\sigma(T)$ not $f_\pi$ which represents the particular case at zero temperature [14]. Therefore, the quantities features will be affected as seen in Fig. 6, where the hedgehog mass reduced to lower values in comparison with Ref. [14]. Another model is called the Nambu Jona-Lasinio model. Osipov et al. [19] consider the eight-quark interactions in the NJL model and investigated the effect of higher-order interactions on the phase transition and the critical temperature only. Hiller et al. [20] consider six-quark interactions in the NJL model and they studied the effect of these interactions on the phase transition and the critical temperature. Therefore, In Refs. [19, 20], the authors did not calculate the nucleon properties at finite temperature in the framework of NJL model.

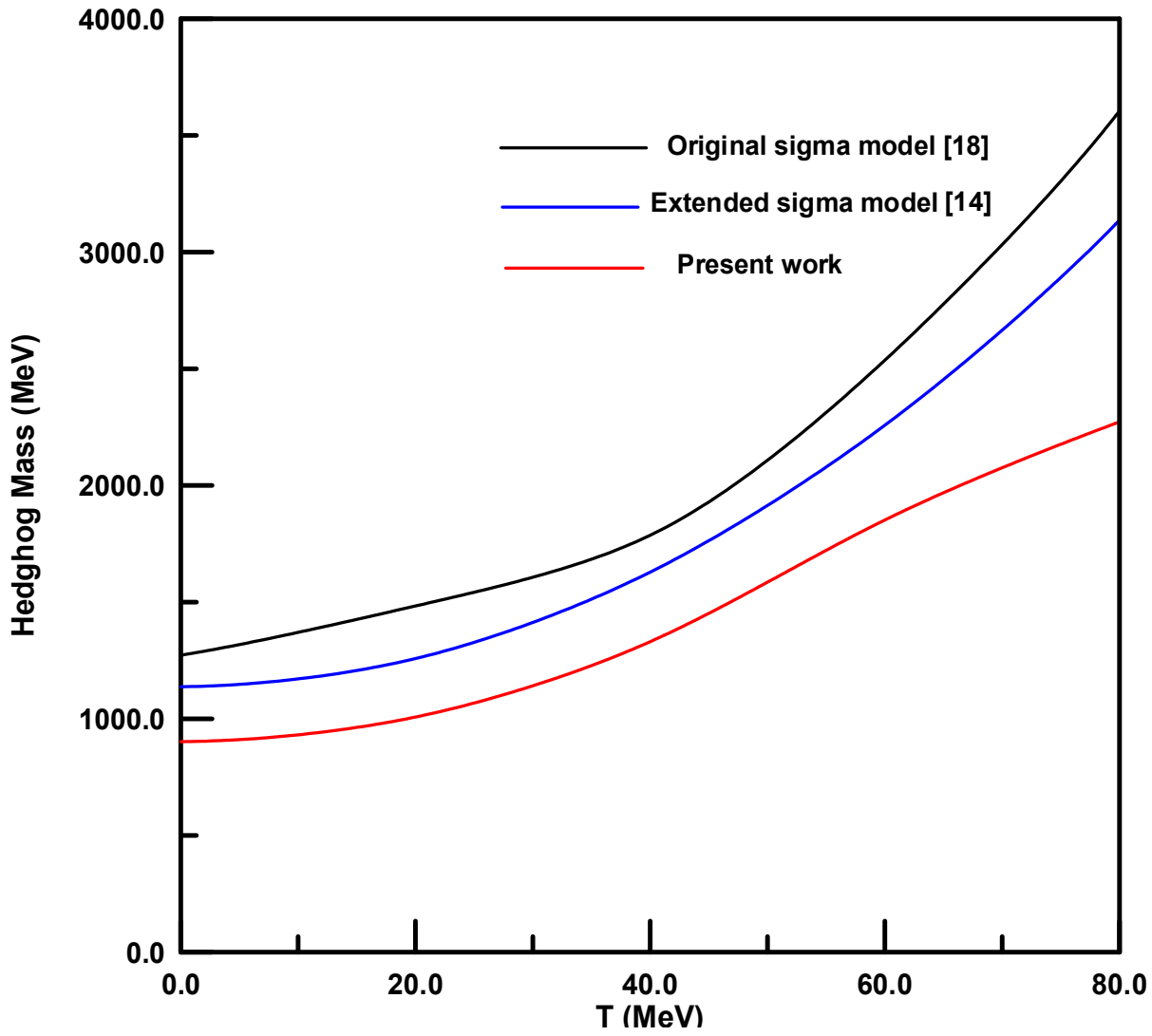

*Fig. 6. The hedgehog mass is plotted as a function of temperature in the original sigma model [18], the extended sigma model [14], and the present work for fixed parameters* $m_\pi = 139.6$ *MeV,* $m_\sigma = 600$ *MeV and* $m_q = 460$ *MeV.*

## 7. Summary and Conclusion

In this work, we extend the previous work [11] to finite temperature. We investigate the effect of finite temperature on the behavior on nucleon properties such the hedgehog mass and the magnetic moments of proton and neutron, and the pion-coupling constant. A comparison with recent works is presented. In addition, we avoid the difficulties which found in the previous works which represent the strength of present model. Despite of the mean-field have been solved by assuming the pion field and quark field take radial direction. This ansatz breaks bath rotational J and isospin I although the grand spin G= I+J is conserved. And also, the mean field describe very light pion flied. Therefore, we need to improve the quality of this work by inserting quantum fluctuations as a future work. In the end, we conclude that the finite temperature with higher-order mesonic interactions play an important role for changing behavior of nucleon properties. In addition, the chiral phase deconfinment is satisfied in the present model.